\documentclass[aps,pra,twocolumn,floatfix]{revtex4-2}
\usepackage{braket}
\usepackage{graphicx}
\usepackage{color}
\usepackage[]{amsmath}
\usepackage{braket}
\usepackage[normalem]{ulem}
\graphicspath{{./images/}}

\begin{document}
\title{Optimal linear optical discrimination of Bell-like states}
\author{Dov Fields$^{1}$, J\'anos A. Bergou$^{2,3}$,  Mark Hillery$^{2,3}$,  Siddhartha Santra$^{1}$, Vladimir S. Malinovsky$^{1}$}
\affiliation{$^{1}$United States Army Research Laboratory, Adelphi, MD 20783\\
$^{2}$Department of Physics and Astronomy, Hunter College of the City University of New York, New York, NY 10065 \\
$^{3}$Graduate Center of the City University of New York, New York, NY 10016}

\begin{abstract}
Quantum information processing using linear optics is challenging due to the limited set of deterministic operations achievable without using complicated resource-intensive methods. While techniques such as the use of ancillary photons can enhance the information processing capabilities of linear optical systems they are technologically demanding. Therefore, determining the constraints posed by linear optics and optimizing linear optical operations for specific tasks under those constraints, without the use of ancillas, can facilitate their potential implementation.  Here, we consider the task of unambiguously discriminating between Bell-like states without the use of ancillary photons. This is a basic problem relevant in diverse settings, for example, in the measurement of the output of an entangling quantum circuit or for entanglement swapping at a quantum repeater station. While it is known that exact Bell states of two qubits can be discriminated with an optimal success probability of 50\% we find, surprisingly, that for Bell-like states the optimal probability can be only 25\%. We analyze a set of Bell-like states in terms of their distinguishability, entanglement as measured by concurrence, and parameters of the beam-splitter network used for unambiguous discrimination. Further, we provide the linear optical configuration comprised of single photon detectors and beam splitters with input state-dependent parameters that achieves optimal discrimination in the Bell-like case. \\
\end{abstract}

\maketitle

\section{Introduction}

Linear optical platforms are a promising route for building quantum information processing devices in computation \cite{kok}, communication \cite{azuma}, and metrology \cite{olson}. On one hand, qubits encoded into the quantum state of a photon can have long coherence times \cite{chen,han} and photonic circuits can potentially be scalably integrated \cite{gan,elshaari,wang,bogaerts}. On the other hand, there are fundamental limitations on the type of operations that can be implemented without prohibitive resource costs. A simple but important example of this kind of limitation is in the case of discriminating measurements on a set of mutually orthogonal entangled pure quantum states. In other platforms, such as superconducting qubits \cite{kjaergaard} or ion traps \cite{bruzewicz}, there are no fundamental limitations on perfectly discriminating between the orthogonal states using measurements in arbitrary orthogonal bases. Whereas, in linear optical systems this is no longer the case: It may not be possible to achieve saturation of the quantum mechanically allowed statistical distinguishability among the given states using only linear optical setups. A case in point being the set of the four maximally entangled states of two qubits, or Bell states, only two of which can be discriminated without the use of ancillary photons.

In principle, given access to certain extra resources such as prepared entangled quantum states and ancillary photons, linear optical elements can be used to implement a universal set of operations for quantum information processing \cite{Knill2001}. In particular, with increasing use of resources, Bell state discriminations can be implemented with a success probability asymptotically approaching $1$ \cite{PhysRevA.84.042331, PhysRevLett.113.140403}. However, increasing the number of ancillary photons to achieve the stated precision is technologically challenging \cite{kok}. Without ancillary photons, only two of the four possible Bell states can be unambiguously discriminated, giving the protocol a maximum efficiency of $50\%$~\cite{Lutkenhaus1999,Vaidman1999,Calsamigilia2001}.

Generalizing this situation is the problem of unambiguously discriminating between a set of mutually orthogonal partially-entangled states of two qubits encoded into four photonic modes, which we call  the set of Bell-like states. The formal structure of Bell-like states in terms of the mode creation operators is identical to that of the Bell states. However, the crucial difference is in the value of their concurrence which is strictly less than 1, i.e., they are partially entangled. Obtaining the linear optical operation that optimally discriminates between Bell-like states is, therefore, an important task since partially entangled states are realistic in the practical scenario. While conditions have been derived in order to determine whether a desired transformation is implementable using linear optics \cite{PhysRevA.69.012302,PhysRevA.73.062320,PhysRevA.100.022301}, these results have limited utility in determining the optimal transformation for specific tasks.

The goal of this paper is to derive the efficiency of optimal linear optical discrimination of Bell-like states and the corresponding setup, i.e., a network of beam-splitters and photon detectors which achieves the optimal efficiency. Our focus is on the case where no ancillary photons are used. The approach is to derive constraints required by unambiguous discrimination between the Bell-like states that allow us to construct feasible linear optical transformations under those constraints. The transformations are then optimized to maximize their probability of success. Completing these steps allows us to design a general method for optimally discriminating any set of Bell-like states. We find that the efficiency, or maximum success probability, of the optimal unambiguous discrimination is only $25\%$, in contrast to the $50\%$ that can be achieved for Bell states~\cite{Lutkenhaus1999,Vaidman1999,Calsamigilia2001}.


The structure of our paper is outlined as follows. In Sec. II, we review the basic mathematical framework underlying linear optical setups for state discrimination. Next, in Sec. III, we define the Bell-like states and proceed to derive the optimal unambiguous discrimination achievable using linear optical setups. We show, in particular, that only two out of the four given states can be successfully discriminated. In Sec. IV, we analyze the optical network allowing the optimal unambiguous discrimination between the Bell-like states showing the  25\%  efficiency of success.  After presenting the results, we conclude by discussing some possible follow-up directions.

\section{Linear Optics Framework for State Discrimination}
Let us consider the discrimination of two-qubit quantum states, employing the dual-rail representation for qubits~\cite{kok,Pryde2010,Lidar2013}. The basic elements of this representation are $m$-mode photons described by the Fock states, $\ket{n_{m}} \equiv \frac{\hat{a}_{m}^{\dag n}}{\sqrt{n!}}\ket{0}$,  where $\hat{a}_{m}^{\dag}$ is the creation operator for the $m$-th photon mode and $n_{m}$ is the number of photons in that mode. Qubit states in the dual-rail representation are given as: $\ket{0} = \ket{1_{1}, 0_{2}} = \hat{a}_{1}^{\dag}\ket{0}, \ket{1} = \ket{0_{1},1_{2}} = \hat{a}_{2}^{\dag}\ket{0}$. Adding a second qubit can be represented
by another photon in two other modes: $\ket{00} = \hat{a}_{1}^{\dag}a_{3}^{\dag}\ket{0}, \ket{01} = \hat{a}_{1}^{\dag}\hat{a}_{4}^{\dag}\ket{0}, \ket{10} = \hat{a}_{2}^{\dag}a_{3}^{\dag}\ket{0}, \ket{11} = \hat{a}_{2}^{\dag}\hat{a}_{4}^{\dag}\ket{0}$. Therefore, the first qubit is represented by one photon in the first two modes and the second qubit is represented by one photon in the second two modes. It is important to note that, by the nature of this representation, the computational space is only a subset of all possible states.

The relevance of this qubit encoding 
is that any transformation allowed by linear optical elements, i.e., any transformation using only beam splitter and phase shifter generators, can be described 
by unitary transformations on the creation and annihilation operators \cite{PhysRevLett.73.58}. We can define the output operators $\left\{ \hat{b}_{i}^{\dag} | i = 1 \dots m \right\}$ in terms of the input operators as $\hat{b}_{i}^{\dag} = \sum_{j} U_{ij}\hat{a}_{j}^{\dag}$.  

\begin{figure}[h]
  \begin{center}
    \includegraphics[scale = 0.7]{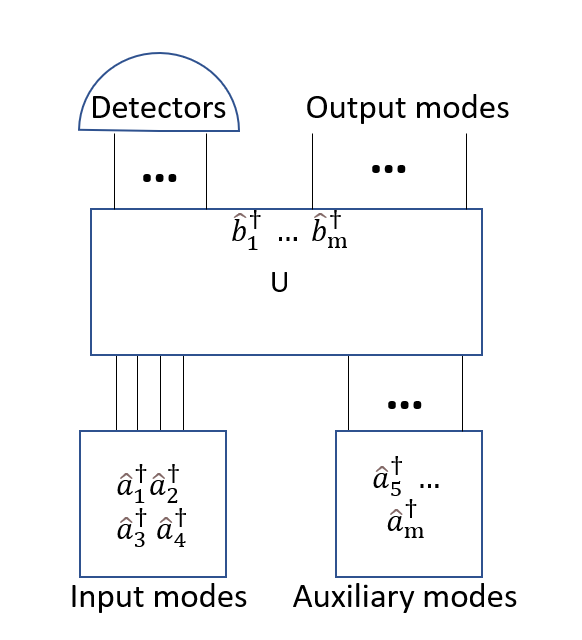}
  \end{center}
  \caption{The general scheme for linear optical operations.  The input modes, $\hat{a}^{\dag}_{i}$ (the system, $i=1, 2, 3, 4$) and the auxiliary modes (the ancilla, $i=5, \dots, m$) are coupled via a network of beam splitters and phase shifters to form the output modes. The action of the linear optic network can be described by a total unitary transformation $U$. At the output, some of the modes are measured using photon resolving detectors, while the remaining undetected modes can be used as input for further processing.
}
  \label{fig:measurement}
\end{figure}

Now we describe the generalized measurement scheme allowed by linear optical setups shown in Fig. \ref{fig:measurement}. At the input of the scheme are the 
photon modes $\left\{ \hat{a}^{\dag}_{i} | i = 1 \dots 4 \right\}$ that can be coupled with auxiliary photon modes $\left\{ \hat{a}^{\dag}_{i} | i = 5 \dots m \right\}$. These input modes are connected to the output modes $\left\{ \hat{b}^{\dag}_{i} | i = 1\dots m \right\}$ utilizing beam splitters and phase shifters.
Some of the output modes can be detected by photon resolving detectors, while the photons in the remaining modes can be treated as a new states that can be used as an input for further processing. 

For the purposes of this paper, we restrict our consideration to a special class of linear optical schemes, where the auxiliary photon modes are empty. Additionally, we focus only on the optimal measurement for a single iteration, barring the use of conditional measurements. 

\section{Bell-like states discrimination}
Bell-like states can be defined as  
\begin{eqnarray}
  \ket{\Psi_{1}} &=& \left( \alpha_{1} \hat{a}_{1}^{\dag}\hat{a}_{3}^{\dag} + \beta_{1} \hat{a}_{2}^{\dag}\hat{a}_{4}^{\dag} \right)\ket{0}, \label{bell1}\\
  \ket{\Psi_{2}} &=& \left( \beta_{1}^{*} \hat{a}_{1}^{\dag}\hat{a}_{3}^{\dag} - \alpha_{1}^{*} \hat{a}_{2}^{\dag}\hat{a}_{4}^{\dag} \right)\ket{0}, \\
  \ket{\Psi_{3}} &=& \left( \alpha_{2} \hat{a}_{1}^{\dag}\hat{a}_{4}^{\dag} + \beta_{2} \hat{a}_{2}^{\dag}\hat{a}_{3}^{\dag} \right)\ket{0}, \\
  \ket{\Psi_{4}} &=& \left( \beta_{2}^{*} \hat{a}_{1}^{\dag}\hat{a}_{4}^{\dag} - \alpha_{2}^{*} \hat{a}_{2}^{\dag}\hat{a}_{3}^{\dag} \right)\ket{0}, 
  \label{belllike}
\end{eqnarray}
where $\alpha_{i}$ and $\beta_{i}$ are the complex coefficients normalized by $|\alpha_{i}|^2 + |\beta_{i}|^2=1$.
The Bell states are recovered for $\alpha_{1} = \beta_{1} = \alpha_{2} = \beta_{2} = 1/\sqrt{2}$. 

As we mentioned above, the most general operation implementable by linear optics 
has the form $\hat{b}_{i}^{\dag} = \sum_{j} U_{ij}\hat{a}_{j}^{\dag}$, 
and the inverse transformation yields $\hat{a}_{i}^{\dag} = \sum_{j}U_{ji}^{*}\hat{b}_{j}^{\dag}$. Therefore we can define an arbitrary state as $\ket{e} = \sum_{j,k \in \sigma} \alpha_{jk}\hat{a}_{j}^{\dag}\hat{a}_{k}^{\dag}\ket{0}$, where $\sigma \equiv \left\{ j,k|j=1,2; k = 3,4 \right\}$. Using these expressions we can derive the relationship between the input 
and the output modes
\begin{eqnarray}
  \ket{e} &=&  \sum_{j,k \in \sigma} \alpha_{jk}\hat{a}_{j}^{\dag}\hat{a}_{k}^{\dag}\ket{0}, \nonumber \\
	&=& \sum_{j,k \in \sigma}\alpha_{jk}\left( \sum_{l} U_{lj}^{*}\hat{b}_{l}^{\dag} \right)\left( \sum_{m} U_{mk}^{*}\hat{b}_{m}^{\dag} \right)\ket{0}, \nonumber\\
	&=& \sum_{m}\sum_{j,k \in \sigma}\alpha_{jk}U_{mj}^{*}U_{mk}^{*}\hat{b}_{m}^{\dag}\hat{b}_{m}^{\dag}\ket{0} \\
	&& + \sum_{l<m, m}\sum_{j,k \in \sigma} \alpha_{jk} \left( U_{lj}^{*}U_{mk}^{*} + U_{mj}^{*}U_{lk}^{*} \right)\hat{b}_{l}^{\dag}\hat{b}_{m}^{\dag}\ket{0}. \nonumber
	\label{outputmodes}
\end{eqnarray}

Since 
the measurements are performed in the orthonormal basis of photon modes, we just need to 
evaluate the probabilities of 
detecting a various 
combination of two photons for a given state. The probability of detecting two photon in mode $m$ is
\begin{eqnarray}
  |\braket{2_{m}|e}|^{2} &=& 2|\sum_{j,k\in \sigma} \alpha_{jk}U_{mj}^{*}U_{mk}^{*}|^{2} = 2|\left(U^{*} N U^{\dag} \right)_{mm}|^{2} \nonumber\\
  &=& 2|\braket{\phi_{m}|N|\phi^{*}_{m}}|^{2} \nonumber\\
  &=&  \frac{1}{2}|\braket{\phi_{m}|\left( N + N^{\top} \right)|\phi^{*}_{m}}|^{2} ,
  \label{probmes2}
\end{eqnarray}
while the detection probability of one photon in mode $m$ and another in mode $n$ is 
\begin{eqnarray}  
  |\braket{1_{m},1_{n}|e}|^{2} &=& |\sum_{j,k \in \sigma} \alpha_{jk}\left( U_{nj}^{*}U_{mk}^{*} + U_{mj}^{*}U_{nk}^{*} \right)|^{2} \nonumber\\
  &=& |(U^{*} N U^{\dag})_{nm} + (U^{*} N U^{\dag})_{mn}|^{2} \nonumber\\
  &=& |\braket{\phi_{n}|\left( N + N^{\top} \right)|\phi^{*}_{m}}|^{2} , 
  \label{probmes11}
\end{eqnarray}
where $N$ is a matrix whose nonzero elements are $N_{jk} \equiv \alpha_{jk}$, 
$\ket{\phi^{*}_{m}}$ is the $m^{th}$ column of $U^{\dag}$, and $\ket{\phi_{m}}$ is the $m^{th}$ column of $U^{\top}$. 

Note that the detection of one photon in mode $m$ and one photon in mode $n$ corresponds to unambiguous discrimination if $ |\braket{1_{m},1_{n}|e}|^{2} = 0$, or $|\braket{2_{m}|e}|^{2} = 0$ if $m =n$,  for three out of the four input states and non-zero for one of them. This detection event 
uniquely identifies the input state. 


Therefore, unambiguous discrimination of an input state enforces constraints on the set of $\left\{\ket{\phi_{m}}\right\}$. If we can identify these sets of constraints,
we can then construct a $U$ matrix that allows for successful discrimination. In order to derive the constraints, it is helpful to note that there is a linear transformation, $\pi$, that maps the vectors $\ket{\Psi_{\mu}}$ to matrices $\pi\left( \ket{\Psi_{\mu}} \right)$ such that $\pi\left( \ket{\Psi_{\mu}} \right) = N_{\mu} + N_{\mu}^{\top}$. In order to understand this transformation, let us define the following matrix $A_{e}$ that is a straightforward transformation of $\ket{e}$:
\begin{eqnarray}
  A_{e} &=& 
  \begin{pmatrix}
    \alpha_{13} & \alpha_{14}\\
    \alpha_{23} & \alpha_{24}
  \end{pmatrix}.
  \label{Amatrix}
\end{eqnarray}
This simple matrix is an element of the 4 dimensional vector space of 2x2 complex matrices with the inner product $tr\left( A^{\dag}B \right)$. We can then explicitly give $\pi\left( \ket{e} \right)$ using this matrix:
\begin{eqnarray}
  \pi\left( \ket{e} \right) &=& 
  \begin{pmatrix}
    0_{2\times2} & A_{e} & 0_{2\times k} \\
    A_{e}^{\top} & 0_{2\times2} & 0_{2 \times k} \\
    0_{k \times 2} & 0 _{k \times 2} & 0_{k\times k} 
  \end{pmatrix}.
  \label{Nmatrix}
\end{eqnarray}
Here, we have defined $0_{k\times j}$ as a matrix of size $k\times j$ with the elements of $0$. Additionally, $k + 4$ is equal to the total number of output modes.  This representation of $\pi\left( \ket{e} \right)$ makes it obvious that $\pi\left( \ket{e} \right)^{\top} = \pi\left( \ket{e} \right)$. In order to see how this operator acts on the $k + 4$ dimensional vector $\ket{\phi_{m}^{*}}$, it is helpful to decompose $\ket{\phi_{m}^{*}}$ as a direct sum of two two-dimensional vectors, $\ket{u_{m}^{*}} \in \mathcal{H}_{2}$ and $\ket{v_{m}^{*}} \in \mathcal{H}_{2}$ and one $k - 4$ dimensional vector, $\ket{w_{m}^{*}} \in \mathcal{H}_{k-4}$: $\ket{\phi_{m}^{*}} \equiv \ket{u_{m}^{*}} \oplus \ket{v_{m}^{*}} \oplus \ket{w_{m}^{*}}$, where:
\begin{eqnarray}
  \ket{u_{m}^{*}} &=& 
  \begin{pmatrix}
    U_{m1}^{*} \\ U_{m2}^{*}
  \end{pmatrix}, \; 
  \ket{v_{m}^{*}} = 
  \begin{pmatrix}
    U_{m3}^{*} \\ U_{m4}^{*}
  \end{pmatrix}, \; 
  \ket{w_{m}^{*}} = 
  \begin{pmatrix}
    U_{m5}^{*} \\ \vdots \\ U_{mk}^{*}  
  \end{pmatrix}. \nonumber \\
\end{eqnarray}
Given this, we can see that $\pi\left( \ket{e} \right)\ket{\phi_{m}^{*}} = A_{e} \ket{v_{m}^{*}} \oplus A_{e}^{\top}\ket{u_{m}^{*}} \oplus 0$.  
\par
Let us now consider the relevant term of the 2 photon detection probabilities for these states: 
\begin{eqnarray*}
  \braket{\phi_{l}|\pi\left( \Psi_{1} \right)|\phi^{*}_{m}} &=& 
  \alpha_{1} \braket{\phi_{l}|\pi\left( \ket{00} \right)|\phi_{m}^{*}} \\
  && + \beta_{1} \braket{\phi_{l}|\pi\left( \ket{11} \right)|\phi_{m}^{*}} \\
    \braket{\phi_{l}|\pi\left( \Psi_{2} \right)|\phi^{*}_{m}} &=& 
    \beta_{1}^{*} \braket{\phi_{l}|\pi\left( \ket{00} \right)|\phi_{m}^{*}} \\
    && - \alpha_{1}^{*} \braket{\phi_{l}|\pi\left( \ket{11} \right)|\phi_{m}^{*}} \\
    \braket{\phi_{l}|\pi\left( \Psi_{3} \right)|\phi^{*}_{m}} &=& 
    \alpha_{2} \braket{\phi_{l}|\pi\left( \ket{01} \right)|\phi_{m}^{*}} \\ 
    &&+ \beta_{2} \braket{\phi_{l}|\pi\left( \ket{10} \right)|\phi_{m}^{*}} \\
    \braket{\phi_{l}|\pi\left( \Psi_{4} \right)|\phi^{*}_{m}} &=& 
    \beta_{2}^{*} \braket{\phi_{l}|\pi\left( \ket{01} \right)|\phi_{m}^{*}} \\
    &&- \alpha_{2}^{*} \braket{\phi_{l}|\pi\left( \ket{10} \right)|\phi_{m}^{*}}
  \label{detectionprobs}
\end{eqnarray*}
If, e.g., we want the detection of photons $l$ and $m$ to contribute to the unambiguous discrimination of $\ket{\Psi_{3}}$, we get the conditions:
\begin{eqnarray}
    \alpha_{1} \braket{\phi_{l}|\pi\left( \ket{00} \right)|\phi_{m}^{*}} &=&  -\beta_{1} \braket{\phi_{l}|\pi\left( \ket{11} \right)|\phi_{m}^{*}} \label{udcond1}\\
  \beta_{1}^{*} \braket{\phi_{l}|\pi\left( \ket{00} \right)|\phi_{m}^{*}} &=&  \alpha_{1}^{*} \braket{\phi_{l}|\pi\left( \ket{11} \right)|\phi_{m}^{*}} \label{udcond2}\\
  \braket{\phi_{l}|\pi\left( \Psi_{3} \right)|\phi^{*}_{m}} &\ne& 0 \label{udcond3} \\
  \beta_{2}^{*} \braket{\phi_{l}|\pi\left( \ket{01} \right)|\phi_{m}^{*}} &=&  \alpha_{2}^{*} \braket{\phi_{l}|\pi\left( \ket{10} \right)|\phi_{m}^{*}}
  \label{udconds}
\end{eqnarray}
The first two conditions can be combined to give:
\begin{eqnarray*}
  &&|\alpha_{1}|^{2}\braket{\phi_{l}|\pi\left( \ket{00} \right)|\phi_{m}^{*}}\braket{\phi_{l}|\pi\left( \ket{11} \right)|\phi_{m}^{*}} \\
  &&= -  |\beta_{1}|^{2}\braket{\phi_{l}|\pi\left( \ket{00} \right)|\phi_{m}^{*}}\braket{\phi_{l}|\pi\left( \ket{11} \right)|\phi_{m}^{*}}.
  \label{udcondtwo}
\end{eqnarray*}
For Bell-like states, $|\alpha_{1}|^{2} \ne |\beta_{1}|^{2}$, so the only way to satisfy this condition and both of Eqs. \eqref{udcond1} and \eqref{udcond2} is for $\braket{\phi_{l}|\pi( \ket{00} )|\phi_{m}^{*}} =\braket{\phi_{l}|\pi( \ket{11})|\phi_{m}^{*}} = 0$. Since $\pi( \ket{11} )$ only has one non-zero singular value, $\braket{\phi_{l}|\pi( \ket{11} )|\phi_{m}^{*}} = 0$ requires that either $\pi(\ket{11} )\ket{\phi_{m}^{*}} = 0$ or $\pi( \ket{11} )\ket{\phi_{l}^{*}} = 0$. If we start by choosing $\pi(\ket{11})\ket{\phi_{m}^{*}}= 0$ we get the following:
\begin{eqnarray*}
  \begin{pmatrix}
    0 & 0 \\
    0 & 1
  \end{pmatrix}\ket{v_{m}^{*}} \oplus 
  \begin{pmatrix}
    0 & 0 \\
    0 & 1
  \end{pmatrix}\ket{u_{m}^{*}} &=& 0 \\
  \ket{v_{m}^{*}} \propto 
  \begin{pmatrix}
    1 \\ 0
  \end{pmatrix} \; \ket{u_{m}^{*}} &\propto&  
  \begin{pmatrix}
    1 \\ 0
  \end{pmatrix} \\
  \ket{\phi_{m}^{*}} = 
  \varphi_{m1}\begin{pmatrix}
    1 \\ 0
  \end{pmatrix} \oplus 
  \varphi_{m2}\begin{pmatrix}
    1 \\ 0
  \end{pmatrix} &\oplus& \varphi_{m3}\ket{w_{m}^{*}}
  \label{11cond}
\end{eqnarray*}
Here, the normalization of $\ket{\phi_{m}^{*}}$ is enforced by the condition $\sum_{i}^{3} |\varphi_{mi}|^{2} = 1$. Additionally, it is worth noting that there exists one alternate solution where both $\ket{v_{m}^{*}} = \ket{u_{m}^{*}} = 0$. However, with such a solution Eq. \eqref{udcond3} cannot be satisfied. Applying the same approach to $\braket{\phi_{l}|\pi\left( \ket{00} \right)|\phi_{m}^{*}} = 0$, we see that choosing $\pi\left( \ket{00} \right)\ket{\phi_{m}^{*}} = 0$ requires $\ket{\phi_{m}^{*}} = 0$. This choice violates the condition in Eq. \eqref{udcond3} and, hence, cannot contribute to unambiguous discrimination. We can conclude from this that any detection of two photons in the same mode, or $m = l$, cannot contribute to unambiguous discrimination. This leaves us with setting $\pi\left( 00 \right)\ket{\phi_{l}^{*}} = 0$, giving: 
\begin{eqnarray}
  \ket{\phi_{l}^{*}} &=& \varphi_{l1}
  \begin{pmatrix}
    0 \\ 1
  \end{pmatrix} \oplus \varphi_{l2} 
  \begin{pmatrix}
    0 \\ 1
  \end{pmatrix} \oplus \varphi_{l3}\ket{w_{l}^{*}}
  \label{10cond}
\end{eqnarray}
The orthogonality condition $\braket{\phi_{l}^{*}|\phi_{m}^{*}} = \delta_{lm}$ is preserved by requiring that $\braket{w_{l}^{*}|w_{m}^{*}} = \delta_{lm}$. With some simple substitution, we can see that the probability of successfully discriminating $\ket{\Psi_{3}}$, given this detection, is $|\braket{\phi_{l}|\pi\left( \psi_{3} \right)|\phi_{m}^{*}}| \propto |\varphi_{l1}\varphi_{m2} + \varphi_{m1}\varphi_{l2}|$. From the normalizations $\sum_{i}^{3} |\varphi_{mi}|^{2} = 1$ and $\sum_{i}^{3} |\varphi_{li}|^{2} = 1$, it is clear that this proportionality term is maximal when $\varphi_{l3} = \varphi_{m3} = 0$. Thus the optimal solution is to reduce our total number of output modes to 4. If we choose $m = 1$ and $l = 2$ we can put all of this together to derive the following unitary:
\begin{eqnarray*}
 &&U^{\dag} = \\ 
 &&\begin{pmatrix}
    \cos \theta_{1}  & 0 & -\sin \theta_{1} e^{-i\varphi_{1}}& 0 \\
    0 & \cos \theta_{2} & 0 & -\sin \theta_{2} e^{-i\varphi_{2}} \\
     \sin\ \theta_{1} e^{i\varphi_{1}} & 0 & \cos \theta_{1}  & 0 \\
    0 & \sin \theta_{2} e^{i\varphi_{2}} & 0 & \cos \theta_{2}  
   
  \end{pmatrix} .
  \label{ucomplete}
\end{eqnarray*}
In this equation, we have satisfied the condition $|\varphi_{i1}|^{2} + |\varphi_{i2}|^{2} = 1 $ by defining $\varphi_{i1} = \cos\theta_{i}$ and $\varphi_{i2} = e^{i\varphi_{i}}\sin\theta_{i}$ for $i = 1,2$.

Using the condition from Eq. \eqref{udconds}, we get that if we want a measurement of $\ket{1_{1},1_{2}}$ to unambiguously discriminate $\ket{\Psi_{3}}$ then we need: 
\begin{eqnarray}
  \beta_{2}^{*}\cos\theta_{1}\sin\theta_{2}e^{-i\varphi_{2}} &=& \alpha_{2}^{*}\cos \theta_{2}\sin\theta_{1}e^{i\varphi_{1}} 
  \label{12cond}
\end{eqnarray}
It is obvious, by inspection, that detections of $\ket{1_{1},1_{3}}$ and $\ket{1_{2},1_{4}}$ cannot unambiguously discriminate any state. Let us now consider a detection in one of the remaining outputs, for instance, $\ket{1_{1},1_{4}}$. This outcome can unambiguously discriminate $\ket{\Psi_{3}}$, but then the system is only succeeding when $\ket{\Psi_{3}}$ is sent. Assuming that we want our detector to be able to succeed for more than one state, we need to look at using this output to discriminate $\ket{\Psi_{4}}$. It is straightforward to show that Eqs. \eqref{udcond1} and \eqref{udcond2} are already satisfied. The only other condition that needs to be satisfied is:
\begin{eqnarray}
  \alpha_{2} \cos\theta_{1}\cos\theta_{2}  &=& \beta_{2} \sin\theta_{1}\sin\theta_{2}e^{i\left(\varphi_{1}+\varphi_{2}\right)}
  \label{23cond}
\end{eqnarray}
Solving these two equations simultaneously gives $\cos\theta_{1} = \sin\theta_{1} = \frac{1}{\sqrt{2}}$ and $\frac{\alpha_{2}}{\beta_{2}} = \tan\theta_{2}e^{i\left( \varphi_{1} + \varphi_{2} \right)}$. Combining this we derive the following unitary:
\begin{eqnarray}
  U^{\dag} &=& 
  \begin{pmatrix}
    \frac{1}{\sqrt{2}} & 0 & -\frac{1}{\sqrt{2}} & 0 \\
    0 & \beta_{2} & 0 & -\alpha_{2} \\
     \frac{1}{\sqrt{2}} & 0 & \frac{1}{\sqrt{2}} & 0 \\
    0 & \alpha_{2}^{*} & 0 & \beta_{2}^{*}    
  \end{pmatrix}
  \label{finalu}
\end{eqnarray}
This unitary will successfully discriminate $\ket{\Psi_{3}}$ and $\ket{\Psi_{4}}$ with an optimal probability of 25\%. While our initial choice of having the $\ket{1_{1},1_{2}}$ mode contribute to unambiguous discrimination was arbitrary, at this point the problem is completely fixed and no other output modes can contribute to unambiguous discrimination. If we chose a different output, for instance $\ket{1_{2},1_{3}}$, we would end up deriving a permutation of the above unitary that also can only succeed in discriminating the states with a 25\% probability. 

\section{Implementation and Analysis}
\begin{figure}[h]
 \begin{center}
    \includegraphics[width = 0.45\textwidth]{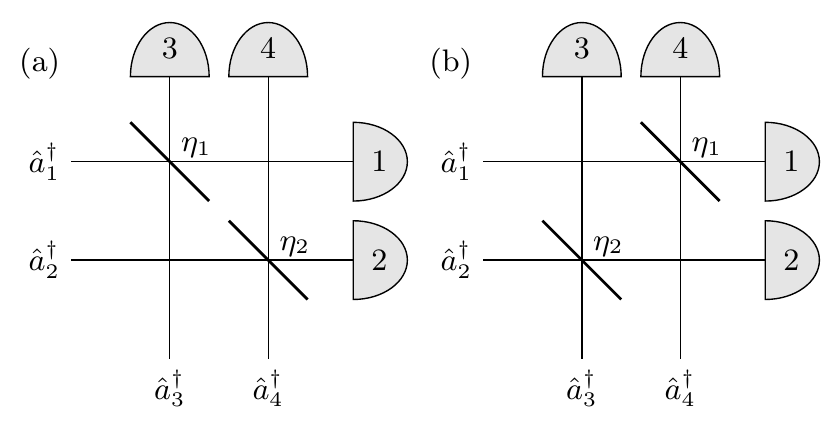}
\end{center}
\caption{Discrimination of Bell-like states, with two beam splitters defined by the parameters $\eta_{1}$ and $\eta_{2}$. In (a), modes 1 and 3 interact in the first beam splitter and modes 2 and 4 interact in the second. In (b), modes 1 and 4 interact in the first beam splitter and modes 2 and 3 interact in the second.}
\label{fig:setup}
\end{figure}

In the previous section, we provide a rigorous proof of the optimal method of discriminating between Bell-like states. In this section, we explicitly provide and analyze the optical setup. Without loss of generality, we can, for convenience, choose all of four parameters in Eqs. \eqref{bell1}-\eqref{belllike} to be real and rewrite the possible Bell-like states in the form:
\begin{eqnarray}
  \ket{\Psi_{1}} &=& \left( \sin \theta_{1} \hat{a}_{1}^{\dag}\hat{a}_{3}^{\dag} + \cos \theta_{1} \hat{a}_{2}^{\dag}\hat{a}_{4}^{\dag} \right)\ket{0}, \\
  \ket{\Psi_{2}} &=& \left( \cos \theta_{1} \hat{a}_{1}^{\dag}\hat{a}_{3}^{\dag} - \sin \theta_{1} \hat{a}_{2}^{\dag}\hat{a}_{4}^{\dag} \right)\ket{0}, \\
  \ket{\Psi_{3}} &=& \left( \sin \theta_{2} \hat{a}_{1}^{\dag}\hat{a}_{4}^{\dag} + \cos \theta_{2} \hat{a}_{2}^{\dag}\hat{a}_{3}^{\dag} \right)\ket{0}, \\
  \ket{\Psi_{4}} &=& \left( \cos \theta_{2} \hat{a}_{1}^{\dag}\hat{a}_{4}^{\dag} - \sin \theta_{2} \hat{a}_{2}^{\dag}\hat{a}_{3}^{\dag} \right)\ket{0}.
  \label{realbelllike}
\end{eqnarray}
We can also easily find the concurrence of these states: $C_{1,2} = \sin\left( 2\theta_{1} \right)$, $C_{3,4} = \sin\left( 2\theta_{2} \right)$. The unitary in Eq. \eqref{finalu} can be implemented by two beam splitters. 
Before looking explicitly at the optimal solution, it is first helpful to consider any general two beam splitter strategy:
\begin{eqnarray}
  \begin{pmatrix}
    \hat{b}_{1}^{\dag}  \\ \hat{b}_{3}^{\dag}
  \end{pmatrix} &=& 
  \begin{pmatrix}
    \eta_{1} & \sqrt{1 - \eta_{1}^{2}} \\
    -\sqrt{1 - \eta_{1}^{2}} & \eta_{1}
  \end{pmatrix}
  \begin{pmatrix}
    \hat{a}_{1}^{\dag} \\ \hat{a}_{3}^{\dag}
  \end{pmatrix} \\
  \begin{pmatrix}
    \hat{b}_{2}^{\dag}  \\ \hat{b}_{4}^{\dag}
  \end{pmatrix} &=& 
  \begin{pmatrix}
    \eta_{2} & \sqrt{1 - \eta_{2}^{2}} \\
    - \sqrt{1 - \eta_{2}^{2}} & \eta_{2}
  \end{pmatrix}
  \begin{pmatrix}
    \hat{a}_{2}^{\dag} \\ \hat{a}_{4}^{\dag}
  \end{pmatrix}
  \label{beamsplitters}
\end{eqnarray}
This interaction is depicted in Fig. \ref{fig:setup} (a), while the setup in Fig. \ref{fig:setup} (b) requires mapping $3 \leftrightarrow 4$ in the previous equations. In order to simplify our analysis, we will fix the first beam splitter to being a 50/50 beam splitter, $\eta_{1} = 1/\sqrt{2}$, and parameterize the second beam splitter by some angle $\phi$, $\eta_{2} = \cos\phi$. 

\begin{table*}[t]
  \caption{Probability of each possible combination of photon detections for each input state.}
  $
\begin{array}{|c|c|c|c|c|c|c|c|c|c|c|}
      \hline
      \text{In/Out} & P(1,1) & P(2,2) & P(3,3)& P(4,4)  & P(1,2) & P(1,3)& P(1,4) & P(2,3) & P(2,4) & P(3,4) \\
      \hline
      \ket{\Psi_{1}} & \frac{\cos^{2}\theta_{1}\sin^{2}\left( 2\phi \right)}{2} & \frac{\sin^{2}\theta_{1}}{2} & \frac{\cos^{2}\theta_{1}\sin^{2}\left( 2\phi \right)}{2} & \frac{\sin^{2}\theta_{1}}{2} & 0 & \cos^{2}\theta_{1}\cos^{2}\left( 2\phi \right) & 0 & 0 & 0 & 0 \\
      \hline
      \ket{\Psi_{2}} & \frac{\sin^{2}\theta_{1}\sin^{2}\left( 2\phi \right)}{2} & \frac{\cos^{2}\theta_{1}}{2} & \frac{\sin^{2}\theta_{1}\sin^{2}\left( 2\phi \right)}{2} & \frac{\cos^{2}\theta_{1}}{2} & 0 & \sin^{2}\theta_{1}\cos^{2}\left( 2\phi \right) & 0 & 0 & 0 & 0 \\
      \hline
      \ket{\Psi_{3}} & 0 & 0 & 0 & 0 &  \frac{\cos^{2}\left( \theta_{2} - \phi \right)}{2} &0 & \frac{\cos^{2}\left( \theta_{2} - \phi\right)}{2} & \frac{\sin^{2}\left( \theta_{2} + \phi \right)}{2} & 0 & \frac{\sin^{2}\left( \theta_{2} + \phi \right)}{2} \\
      \hline
      \ket{\Psi_{4}} & 0 & 0 & 0 & 0 & \frac{\sin^{2}\left( \theta_{2} - \phi \right)}{2} & 0 & \frac{\sin^{2}\left( \theta_{2} - \phi \right)}{2} & \frac{\cos^{2}\left( \theta_{2} + \phi \right)}{2} & 0 & \frac{\cos^{2}\left( \theta_{2} + \phi \right)}{2} \\
      \hline
\end{array}
$
\label{tab:probtable}
\end{table*}

Using this, we calculate the probabilities of measuring each possible outcome for each possible input state.
In the table~\ref{tab:probtable}, $P(m,n)$ is the probability of detecting one photon in detector $m$ and one photon in detector $n$. If we swap the two beam splitters, then we get the same table, but with $1 \leftrightarrow 2$ and $3 \leftrightarrow 4$. If we swap the interactions, as depicted in Fig. \ref{fig:setup} (b),  such that the first beam splitter has the $1$ and $4$ modes as its input and the second beam splitter has $2$ and $3$ as it's input, we also get a similar table, but with the first two and the last two rows of this table swapped and with $\theta_{1} \leftrightarrow \theta_{2}$. For each output, we can use Bayes' Theorem to calculate the confidence \cite{PhysRevLett.96.070401,PhysRevA.79.032323}:
\begin{eqnarray}
  P\left(\ket{\psi_{i}}|m,n \right) &=& \frac{P(m,n|\ket{\psi_{i}})p\left( \ket{\psi_{i}} \right)}{\sum_{i}P\left(m,n|\ket{\psi_{i}}\right)p\left( \ket{\psi_{i}} \right)} \label{bayes} \\
  D\left( m,n \right) &=& \max_{i}\left\{ P\left( \ket{\psi_{i}}|m,n\right) \right\}
  \label{dist}
\end{eqnarray}
Here, we have defined $P\left( m,n|\ket{\psi_{i}} \right)$ as the probability of a given detection outcome of two photons in the  $m$ and $n$ detectors for the input state $\ket{\psi_{i}}$. In addition, we have assumed that all Bell-like states are sent with equal probability: $p\left( \ket{\psi_{i}}\right) = \frac{1}{4}$. One final note is that Eq. \eqref{bayes} only holds when the denominator is non-zero. This definition of confidence gives a measure of how well a given detection can be correlated to one of the input states. When the confidence is $\frac{1}{4}$ there is no correlation between the detection and any input state and when the confidence is 1 there is perfect correlation between the detection and the associated input state. In the case where confidence is 1, that detection can contribute to unambiguous discrimination. From all the columns of the table we only get 3 different equations for confidence:
\begin{eqnarray}
  D\left( 1,1 \right) &=&  D\left( 2,2 \right) = D(3,3) = D(4,4) = D\left( 1,3 \right) \equiv D_{1} \nonumber\\
  D\left( 1,2 \right) &=& D\left( 1,4 \right) \equiv D_{2} \nonumber\\
  D\left( 2,3 \right) &=& D\left( 3,4 \right) \equiv D_{3} \nonumber\\
  D_{1} &=& \frac{1 + \sqrt{1 - C_{1}^{2}}}{2} \\
  D_{2} &=& \frac{1 + |\sqrt{1 - C_{2}^{2}}\cos\left(2\phi\right) + C_{2}\sin\left( 2\phi \right)|}{2} \\
  D_{3} &=& \frac{1 + |\sqrt{1 - C_{2}^{2}} \cos\left( 2\phi \right) - C_{2}\sin\left( 2\phi \right)|}{2}
  \label{distcalc}
\end{eqnarray}  

\begin{figure}[h]
  \begin{center}
    (a)
    \includegraphics[width = 0.45\textwidth]{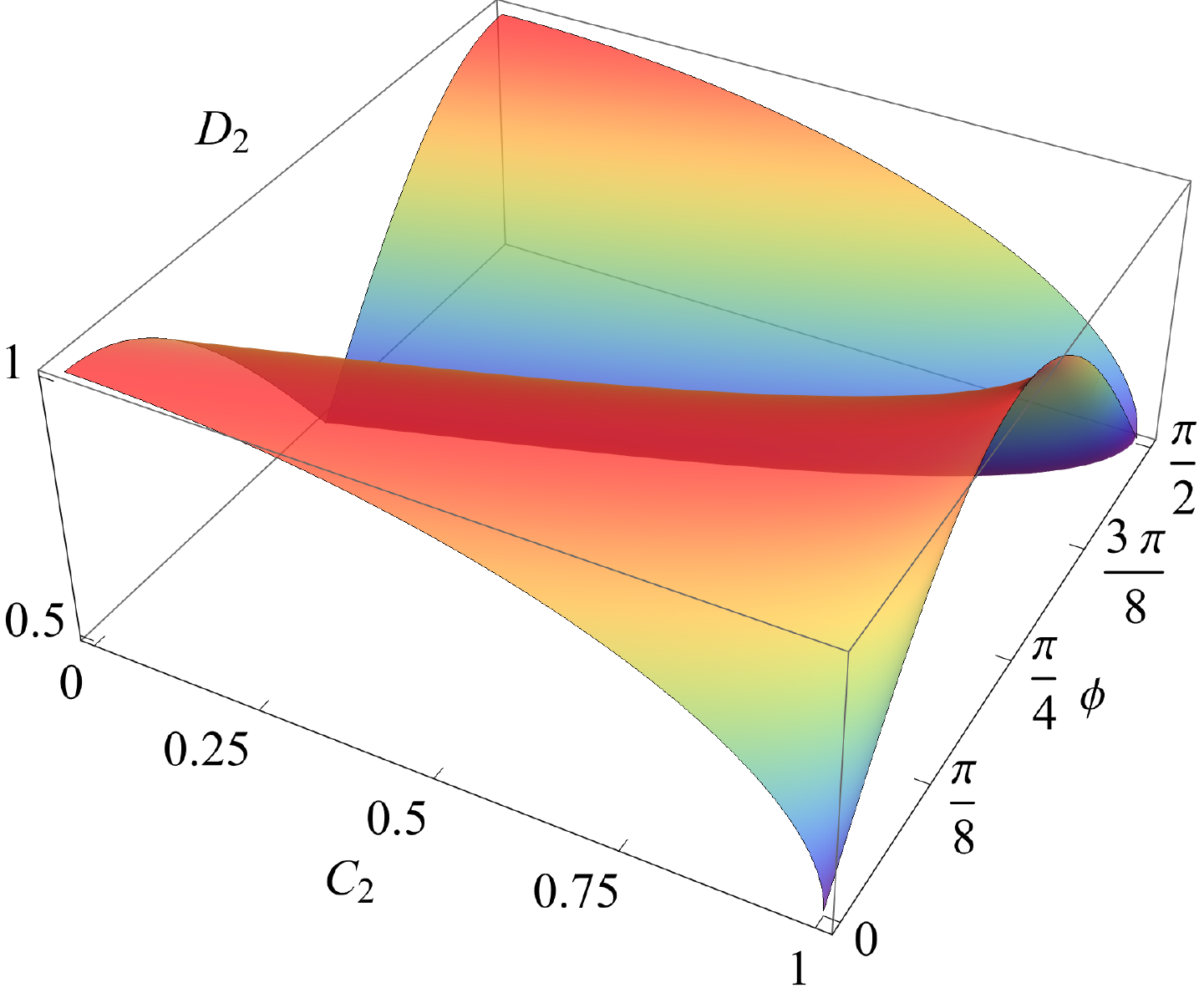}
    (b)
    \includegraphics[width = 0.45\textwidth]{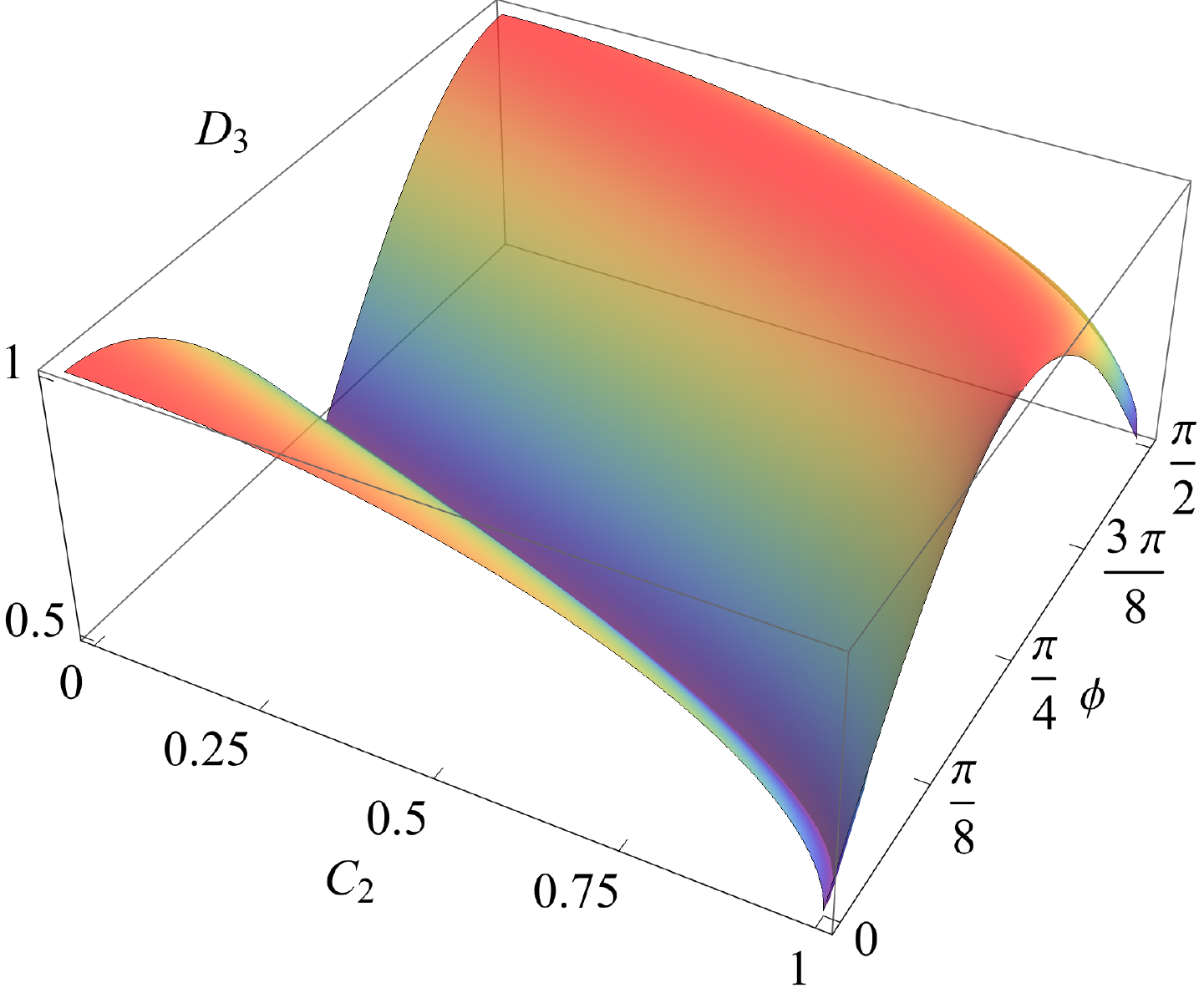}
  \end{center}
  \caption{A plot of the confidences (a) $D_{2}$ and (b) $D_{3}$ as a function of the concurrence $C_{2}$ and the beam splitter parameter $\phi$.}
  \label{fig:dist}
\end{figure}

In Fig. \ref{fig:dist}, we illustrate a plot of both $D_{2}$ and $D_{3}$ as functions of $C_{2}$ and $\phi$. Since unambiguous discrimination is only achieved when the confidence is 1, we can see that $D_{1}$ only contributes to unambiguous discrimination when the first two states are separable, or $C_{1} = 0$. In order to satisfy either $D_{2} = 1$ or $D_{3} = 1$, we only need to choose $\phi = \theta_{2}$ or $\phi = \frac{\pi}{2} -\theta_{2}$ respectively, which is the optimal solution derived in the previous section and results in the unitary given in Eq. \eqref{finalu} up to a simple permutation. For both $D_{2} = 1$ and $D_{3} = 1$ to be satisfied, we either need $C_{2} = 1$ and $\phi = \frac{\pi}{4}$, which is the case for Bell states, or $C_{2} = 0$ and $\phi = 0$, which is the case for separable states. This analysis makes it clear that the unambiguous linear optical discrimination of Bell-like states is not a monotonic function of entanglement, or equivalently, concurrence. Rather, for all Bell-like states, only one of $D_{2} = 1$ or $D_{3} = 1$ can be satisfied, and therefore the Bell-like states can only be successfully discriminated with a probability of 25\%. For Bell states, both $D_{2} = 1$ and $D_{3} = 1$ can be satisfied, allowing for a success probability of 50\% for the discrimination. For completely separable states, $D_{1} = D_{2} = D_{3} = 1$ can be satisfied, allowing for complete discrimination between the four states.

\section{Conclusion}
In this paper we have derived the optimal efficiency of unambiguous discrimination between Bell-like states possible with linear optical setups without the need for ancillary photons. We have explicitly shown that the optimal efficiency for Bell-like states is only 25\%, as opposed to the 50\% success rate possible for Bell states. The reduced symmetry of the Bell-like states results in fewer outputs that can be useful for unambiguous discrimination. When analyzed in terms of the entanglement measure of the set of states, the optimal efficiency shows a discontinuity between the set of Bell-like states and exact Bell-states. The main conclusion is that the upper bound for the success probability of unambiguous discrimination between Bell-like states is 25\%. This result is independent of the concurrence $C$ of the states for $0<C<1$, while $C = 0$, separable states, and $C=1$, maximally entangled states, emerge as singular points. Previous works on Bell states simply prove that the proposed transformation is optimal, in this paper we obtained specific constraints on the unitary and used these constraints to derive and construct the optimal discrimination protocol. The systematic approach presented in this paper has the potential to assist in optimizing other types of linear optical discrimination problems. In follow up work, we intend to consider more general classes of orthogonal entangled states. In addition, there is still room to explore optical setups for unambiguous discrimination that make use of ancillary photons. One final possible extension of this work is using this approach to derive the optimal minimum error discrimination, or even more general strategies, allowed by linear optical setups.

\section{Acknowledgements}
This research was sponsored by the Army Research Laboratory and was accomplished under Cooperative Agreement Number W911NF-20-2-0097. The authors would like to thank Michael Goerz for helpful discussions.


\begin{thebibliography}{99}
\bibitem{kok} P. Kok, W. J. Munro, Kae Nemoto, T. C. Ralph, J. P. Dowling, and G. J. Milburn, \rmp {\bf 79}, 135 (2007).
\bibitem{azuma} K. Azuma, K. Tamaki, and H. K. Lo, Nat. Commun {\bf 6}, 6787 (2015). 
\bibitem{olson} J. P. Olson, K. R. Motes, P. M. Birchall, N. M. Studer, Margarite LaBorde, T. Moulder, P. P. Rohde, and J. P. Dowling, \pra {\bf 96}, 013810 (2017).
\bibitem{chen} J. F. Chen, Z. Han, P. Qian, L. Zhou, and W. Zhang, in Proc. 11th Conference on Lasers and Electro-Optics Pacific Rim (CLEO-PR), 2015, pp. 1-2.
\bibitem{han} Z. Han, P. Qian, L. Zhou, {\it{et al.}}, Sci Rep {\bf 5}, 9126 (2015). 
\bibitem{gan} X. Gan, R. J. Shiue, Y. Gao, I. Meric, T. F. Heinz, K. Shepard, J. Hone, S. Assefa, {\it{et al.}}, Nat. Photonics {\bf 7}, 883 (2013).
\bibitem{elshaari} A. W. Elshaari, W. Pernice, K. Srinivasan, {\it{et al.}}, Nat. Photonics {\bf 14}, 285 (2020). 
\bibitem{wang} J. Wang, F. Sciarrino, A. Laing, {\it{et al.}}, Nat. Photonics {\bf 14}, 273 (2020).
\bibitem{bogaerts} W. Bogaerts, D. P\'{e}rez, J. Capmany, {\it{et al.}}, Nature {\bf 586}, 207 (2020). 
\bibitem{kjaergaard} M. Kjaergaard, Mollie E. Schwartz, J. Braum\"{u}ller, P. Krantz, Joel I.-J. Wang, S. Gustavsson, and W. D. Oliver, Annual Review of Condensed Matter Physics {\bf 11}, 369 (2020).
\bibitem{bruzewicz} C. D. Bruzewicz, J. Chiaverini, R. McConnell, and J. M. Sage, Applied Physics Reviews {\bf 6}, 021314 (2019).
\bibitem{Knill2001} E. Knill, R. Laflamme, and G. J. Milburn, Nature  {\bf 409}, 46 (2001).
\bibitem{PhysRevA.84.042331} W. P. Grice, \pra {\bf 84}, 042331 (2011).
\bibitem{PhysRevLett.113.140403} F. Ewert and P. van Loock, \prl {\bf 113}, 140403 (2014).
\bibitem{Lutkenhaus1999} N. L\"{u}tkenhaus, J. Calsamiglia, and K.-A. Suominen, \pra {\bf 59}, 3295 (1999). 
\bibitem{Vaidman1999} L. Vaidman and N. Yoran, \pra {\bf 59}, 116 (1999). 
\bibitem{Calsamigilia2001} J. Calsamiglia and N. L\"{u}tkenhaus, Applied Physics B {\bf 71}, 67 (2001).
\bibitem{PhysRevA.69.012302} P. van Loock and N. L\"{u}tkenhaus, \pra {\bf 69}, 012302 (2004). 
\bibitem{PhysRevA.73.062320} P. van Loock, K. Nemoto, W. J. Munro, P. Raynal, and N. L\"{u}tkenhaus, \pra {\bf 73}, 062320 (2006).
\bibitem{PhysRevA.100.022301} J. C. Garcia-Escartin, V. Gimeno, and J. J. Moyano-Fern\`{a}ndez, \pra {\bf 100}, 022301 (2019),
\bibitem{Pryde2010}
T. C. Ralph and G. J. Pryde, Progress in Optics {\bf 54}, 209 (2010).
\bibitem{Lidar2013} 
L. A. Wu, P. Walther, D. Lidar, Sci. Rep. {\bf 3}, 1394 (2013). 
\bibitem{PhysRevLett.73.58} M. Reck, A. Zeilinger, H. J. Bernstein, and P. Bertani, \prl {\bf 73}, 58 (1994)
\bibitem{PhysRevLett.96.070401}S. Croke, E. Andersson, S. M. Barnett, C. R. Gibson, end J. Jeffers, \prl {\bf 96}, 070401 (2006).
\bibitem{PhysRevA.79.032323}U. Herzog, \pra {\bf 79}, 032323 (2009).
\end{thebibliography}
\end{document}